\documentstyle[preprint,prd,aps,floats,tighten,epsf]{revtex}
\begin{document}
\def\bull{\ {\vrule height 1.4ex width 1.4ex depth 0.2ex} }
\preprint{\vbox{\hbox{RUHN-00-02}}}

\title{Exact local fermionic zero modes}

\author{Federico Berruto\footnote{
On leave from the Dipartimento di Fisica and Sezione INFN,
University of Perugia, Via Pascoli I-06123, Perugia, Italy.}}
\address{Department of Physics and Astronomy,
Rutgers University, Piscataway, NJ 08855-0849
{\tt berruto@physics.rutgers.edu}}
\author{Rajamani Narayanan}
\address{American Physical Society,
One Research Road, Ridge, NY 11961\\ {\tt rajamani@bnl.gov}}
\author{Herbert Neuberger}
\address{Department of Physics and Astronomy,
Rutgers University, Piscataway, NJ 08855-0849
\\ {\tt neuberg@physics.rutgers.edu}}

\maketitle
\begin{abstract}
We introduce a simple method to find localized exact fermionic zero modes
for any local fermionic action. The zero modes are attached to specific
local gauge configurations. Examples 
are provided for staggered and Wilson fermion actions in 2-6  
dimensions, at finite and infinite lattice volumes, and for abelian and
non-abelian gauge groups. One of our concrete results is that a finite density
of almost zero modes must occur in quenched four dimensional 
lattice gauge theory simulations that use traditional methods.  
This density is exponentially suppressed in the gauge
coupling constant.  
\end{abstract}

\section{Introduction}

Lattice fermions in interaction with background gauge fields have a tendency
to produce localized fermionic approximate zero modes. Since the zero
modes are localized, the influence of gauge fields far away from the center
of the zero mode is small and should be well approximated by a mean-field,
uniform configuration. One would guess then that what causes the zero mode
is a certain local structure in the gauge configuration. 
In this paper we show how one
can easily test whether a given local gauge configuration will have an
exact zero mode at finite or infinite lattice size. We also show that one
can get approximate zero modes in this way, using the variational principle.
Inverting the logic, we show how one can find local gauge structures that
bind fermionic (approximate) zero modes to them.

There are two main classes of fermionic 
lattice zero modes: relatively large ones
that should scale and have a physical effect in the continuum limit 
and very local ones, whose size shrinks to zero on continuum scales 
and therefore are lattice artifacts. 
The lattice artifact zero modes
are a serious impediment to practical simulations in the quenched
approximation~\cite{EHN} and to the implementation
of exact chirality on the lattice using 
the overlap Dirac operator ~\cite{overlap}. 

A better understanding of the
fermionic zero modes is important in both cases: we want to see their
effects clearly when these are genuine continuum effects ~\cite{ovpinst}
and would like
to suppress them when they are lattice artifacts.
There is a difference between the roles the approximate
zero modes which are lattice artifacts play in the traditional
context and in the overlap context. In a traditional quenched
QCD calculation the lattice artifacts can become an unsurmountable problem
because their effect on fermion 
propagators can be spuriously large and there is no natural
prescription for how to handle these non-universal effects. 
In the context of the overlap
however, the problem is less severe because there is a well defined 
natural procedure to deal with these lattice artifacts. The problem
is now only of a numerical nature, since the evaluation of the
sign function for numerically tiny arguments is computationally
costly. Strictly speaking, {\it exact} zero modes are not a 
major difficulty in either framework. Still, when trying to understand
the origin of the approximate fermionic zero modes, it is best to
start analytical work from exact ones.

\section{The basic idea}

Let $S$ denote a cluster comprising of
a finite local collection of sites $s$ and let the gauge field
configuration have $U_\mu(x) \equiv 1$ on all links outside the cluster.
Let $V_S$ be the subspace of the space $V$ of fermion fields $\psi$ 
with $\psi(x)=0$ for $x\notin S$. Let $D_f$ be the appropriate sparse
matrix realization of the free lattice Dirac operator with uniform
link variables $U_\mu (x) \equiv 1$. Let us assume that $G_f=D_f^{-1}$
exists. The total Dirac operator, $D$, maps $V$ into $V$ and
is given by
\begin{equation}
D=L+D_f
\end{equation}
$L$ maps the entire $V$ into $V_S$.
 
Consider now the operator 
\begin{equation}
R=1+LG_f
\end{equation} 
It naturally defines a restricted operator $R_S$ from $V_S$ to $V_S$. 

The equation $D\psi =0$ has a solution in $V$ if and only if there is
a $\phi\in V_S$ such that $R_S \phi =0$. (At infinite volume this
solution will be normalizable if the decay of $G_f$
is fast enough.) The main point is that $R_S$ is a small matrix if
the cluster is small. 

The proof of the above assertion is easy:
If $\psi$ is a zero mode of $D$ define $\phi = L \psi$; clearly,
$\phi \in V_S$ and $L(1+G_f L)\psi =0$. If $\phi \in V_S$ obeys
$R_S \phi =0$, define $\psi=-G_f \phi$. Since $\phi = -L G_f\phi$,
we also have $L\psi=\phi$ implying $D\psi =0$.  

The above observations can be easily extended to the case that
$R_S^\dagger R_S$ has a small eigenvalue $\eta$. Then, one gets a variational
upper bound for the lowest eigenvalue of $D^\dagger D$. If there are
several linearly independent small eigenvalue eigenstates of $R_S$
then we have bounds on several of the lowest eigenvalues of $D^\dagger D$.
For a single state $\phi \in V_S$ satisfying
\begin{equation}
R_S^\dagger R_S \phi = \eta \phi
\end{equation}
we make the variational ansatz
\begin{equation}
\psi=-G_f \phi
\end{equation}
and find
\begin{equation}
\lambda_{\rm min} (D^\dagger D )\le \eta {{\phi^\dagger \phi}\over
{\phi^\dagger {1 \over{D_f^\dagger D_f}} \phi }}
\end{equation}

\section{``Fluxon'' configuration with staggered fermions}

This is an example that was partially analyzed before~\cite{oldnpb}. 
The gauge configuration consists of making all links that go out into
the positive directions of a fixed site be $-1$. 
All other links are set to unity. This configuration was called a ``fluxon''
in the past. It is a local minimum of the single plaquette 
action with gauge groups $U(1)$ or $SU(2)$ and with both a fundamental
and an adjoint term. For $SU(2)$ it is known 
that there is a crossover along the
Wilson line which is related to an end point of a transition line
in the extended fundamental-adjoint plane~\cite{bhancreu}. 
The transition across that
line can be argued to be related to a condensation of fluxons. At
weak coupling there is a finite and calculable density of local
fluxons. 

We first consider naive fermions in a single fluxon background. In $d$
dimensions we shall find, at infinite volume, several degenerate normalizable 
fermionic zero modes. The decay of the zero modes is power-like, since
the fermionic theory is massless.

In $d$ dimensions the $\gamma_\mu$ matrices
are $2^{d_2} \times 2^{d_2}$, where $d_2$ is the integer
part of ${d\over 2}$.  
A simple calculation produces the following $(d+1) 2^{d_2} \times
(d+1) 2^{d_2} $ $R_S$ matrix:
\begin{equation}
R_S =\pmatrix{ 1 -{1\over d} \gamma \gamma^T & 0\cr
                       0 & 0\cr}
\end{equation}
Here we introduced a $2^{d_2}d\times 2^{d_2}$ matrix $\gamma$ and
the $2^{d_2}\times 2^{d_2}d$ matrix $\gamma^T$:
\begin{equation}
\gamma=\pmatrix {\gamma_1\cr\gamma_2\cr\vdots\cr\gamma_d\cr}, ~~~
\gamma^T=\pmatrix {\gamma_1 &\gamma_2 & \dots &\gamma_d\cr}
\end{equation}

The most evident zero mode (normalizable for $d\ge 3$) is given by
choosing 
\begin{equation}
\phi=\pmatrix{0\cr 0\cr\vdots\cr\chi\cr}
\end{equation}
Each entry above has $2^{d_2}$ components. We then find $2^{d_2}-$
fold degenerate zero modes given by (up to normalization factors):
\begin{equation}
\psi_0 (x) =-\int{{d^d p}\over {(2\pi )^d }}  {
{\sin px }\over {\sum_\mu \gamma_\mu
\sin p_\mu }} \chi
\end{equation}
This class of zero modes has been found previously in~\cite{oldnpb}. 
Our more detailed study here shows 
that there are more zero modes: Choose $\phi^\prime$ 
\begin{equation}
\phi^\prime = \pmatrix {\gamma_1 \chi^\prime\cr
\gamma_2 \chi^\prime\cr\vdots\cr\gamma_d \chi^\prime \cr 0 }
\end{equation}
It is easy to check that 
\begin{equation}
(1-{1\over d} \gamma\gamma^T ) \phi^\prime =0
\end{equation}
The new zero modes of $D$ are
\begin{equation}
\psi_0^\prime (x) = i\delta_{x,0} \chi^\prime +
{1\over 2} \int{{d^d p}\over {(2\pi )^d }} 
\sin px  {1\over {\sum_\mu \sin^2 p_\mu }} \left [
\sum_{\mu\ne\nu} \gamma_\mu\gamma_\nu \sin (p_\mu - p_\nu ) +
\sum_\mu \sin 2p_\mu \right ] \chi^\prime
\end{equation}

In four dimensions for example, we have found 8 zero modes. There are
no other zero modes. Each set
of four zero modes maps into itself under the action of $\gamma_5$.
One can easily decompose each set of 4 zero modes into 2 of positive
chirality and 2 of negative chirality. 
One naive lattice fermion
should represent 16 continuum fermions. The fluxon configuration background
breaks the continuum flavor symmetry because there are half as many
zero modes as expected continuum fermions. 

Consider now staggered fermions; they are treated by block diagonalizing
the naive fermion system. 
When we decouple the four dimensional naive fermions into
four staggered fermions, each will have two zero modes. In the continuum
limit each staggered fermion should reproduce four flavors. With only
two zero modes for a staggered fermion it is again clear that the fluxons 
are nonperturbative contributors to lattice flavor symmetry breaking.

\section{``Fluxon'' configuration with Wilson fermions}

The zero modes we found above were localized only by power decays.
To get light quarks with Wilson fermions the mass parameter has
to be given negative values. The free propagator is now massive,
decaying exponentially. We use the same technique to look for negative
mass values where fluxons can produce exact, exponentially localized,
fermionic zero modes. 

We work in dimension $d$ larger or equal to 2. 
The matrix $R_S$ now becomes:
\begin{equation}
R_S = 1 + \pmatrix {(1+\gamma_1 ) G_f (0,{\hat 1})& \dots& 
(1+\gamma_1 )  G_f (0,{\hat d}) & (1+\gamma_1 )  G_f (0,0)\cr
\vdots&\vdots &\vdots &\vdots \cr
(1+\gamma_d ) G_f (0,{\hat 1})& \dots& 
(1+\gamma_d )  G_f (0,{\hat d}) & (1+\gamma_d )  G_f (0,0)\cr
\sum_\mu (1-\gamma_\mu ) G_f ({\hat \mu},{\hat 1)})&
\dots & \sum_\mu (1-\gamma_\mu ) G_f ({\hat \mu},{\hat d})&
\sum_\mu (1-\gamma_\mu ) G_f ({\hat \mu},0)}
\end{equation}

To search for zero modes we calculate the determinant of $R_S$.
It inherits reality from the determinant of $D$. By working
out explicitly the cases $d=2,3,4,5,6$ we arrive at a formula for general $d$.
We have not derived the formula for arbitrary $d$, so we present it
as a conjecture.  
\begin{equation}
\det R_S = 
\left [ 1+{{d-1}\over d} [x(m+d)-1]^2 + 2[x(m+d)-1] + d(d-1)xy \right ]^{2^{d_2}}
\end{equation}
The parameters $x$ and $y$ are defined by fermionic propagators: 
\begin{eqnarray}
x=\int{{d^d p}\over {(2\pi )^d }} {b(p)\over {b^2 (p) + s^2 (p) }}\quad &,&
~~~
y=\int{{d^d p}\over {(2\pi )^d }}{{\cos p_1 [ 2\sin^2 p_2 - b(p) \cos p_2 ]}
\over{b^2(p) +s^2 (p) }}\quad ,
\nonumber\\
b(p)= m+d -\sum_\mu \cos p_\mu \quad &,&
~~~ s^2(p)=\sum_\mu \sin^2 p_\mu \quad .
\nonumber\\
\end{eqnarray}
When there are zero modes, they will be $2^{d_2}$-fold degenerate (at least
for $d=2,3,4,5,6$). 

It is a simple matter now to search for an $m$ for which there are
zero modes. We start from $m=0$ and search going towards negative
values. The first mass values which gives zero modes is listed below.
As far as simulations go, only the region $-2 < m <0$ is of interest.
For $d=4$ the first zero is at $m=-2.87982$, for $d=3$ the first zero
is at $m=-1.940553$, and for $d=2$ the first zero is at $m=-.90096$. 
So, in four dimensions the fluxon is not of great importance, but
in two dimensions it is. Note that the fermions do not react to
the fluxon as they would to an instanton in the continuum as reflected
by the zero-mode degeneracy we find. 

\section{A small two dimensional instanton}
We now turn to analyze a small two dimensional instanton. It is
described by four plaquettes making up a square, with each plaquette
carrying ${\pi\over 2}$ units of angular flux. 

The calculations are now more involved, but the bottom line is that
we find a non-degenerate zero at $m=-0.39182$, which certainly is of
interest. For this value of $m$ the free propagator decays quite fast
so the zero mode is quite local. Effects of boundary conditions can
be easily investigated by appropriately adjusting
$D_f$ and the gauge background. They are more sizable with periodic boundary
conditions than with antiperiodic boundary conditions. We have also
looked at the dependence on the toron coordinates (zero Fourier modes
of the gauge fields $A_\mu (x)$ where the link variables are $U_\mu(x)=
e^{iA_\mu (x)})$. Boundary conditions are implemented by choosing 
specific toron configurations. Finite volume calculations require the 
replacement of the momentum integrals by the appropriate sums. 

\section{Finding gauge configurations that bind zero modes}

Choosing a cluster of sites of certain shape we now view the
link variables residing on the links in the cluster as unknowns.
For each such gauge field configuration we numerically compute
the smallest eigenvalue of $R^\dagger_S R_S$, 
$\lambda_{\rm min} (R^\dagger_S R_S ) $, using Wilson
fermions at some negative value of the mass parameter
$m$. Varying the gauge background, and proceeding by steepest descent
we can find local minima of $\lambda_{\rm min} (R^\dagger_S R_S ) $
as a function of the link variables. Often, we end up
finding the minimum to be zero and 
then the corresponding gauge field configuration binds a 
localized exact zero mode. This search for a gauge configuration can
be repeated for different starting gauge configurations and different
values of the mass parameter. To speed the numerical procedure up
we took the 
free propagators appearing in $R_S$ on a finite
periodic lattice of size $8^d$.

Restricting ourselves to $U(1)$ fields, in 3D, on a cluster
comprising of four sites that can support a fluxon, we find
zero modes only for $m< -1.9$. The minimizing configuration is
not a fluxon and it depends upon the gauge configuration we start
our steepest descent search from. Nevertheless, always, 
the Wilson plaquette
action of the minimizing configuration is close to the one
of a fluxon. As long as we restrict our gauge configurations
to having nontrivial links only along the links contained in the cluster,
different link configurations are gauge inequivalent. 
Thus, there are many gauge inequivalent configurations
that can produce fermionic zero modes. 

In four dimensions the situation is similar for the fluxon type cluster
but now 
we find zero modes only for $m < -2.5$, which is outside of the
range of immediate interest.  

Our next step is to see what happens when we increase the
cluster size. We consider now larger clusters that make
up a cube in 3D or a hypercube in
4D. Keeping the gauge fields still in $U(1)$ we find that now we
can get zero modes at masses closer to zero. In 3D zero modes
are occurring for $m < -1.1 $ and in 4D for $m < -1.7$. Thus,
a moderate increase in cluster size increases the chances to
produce zero modes at Wilson masses relevant to $QCD$ simulations. 

Finally we also look at what happens when one goes from
the abelian $U(1)$ gauge group to the non-abelian 
$SU(2)$ gauge group. We find
that making the non-abelian gauge group always allows certain
gauge configuration to bind zero modes for Wilson masses closer
to zero than in the abelian case. Thus, increasing the size of the group
also increases the chances to produce zero modes at practically
relevant values of the Wilson mass parameter $m$.

\begin{figure}
\epsfxsize = 0.8\textwidth
\caption{ 
Plot of $\lambda_{\rm min}(R_S^\dagger R_S)$ as a function of mass
for four different SU(2) clusters. Two of them are in 3D
and two of them are in 4D.
}
\centerline{{\setlength{\epsfxsize}{6in}\epsfbox[20 20 600 600]{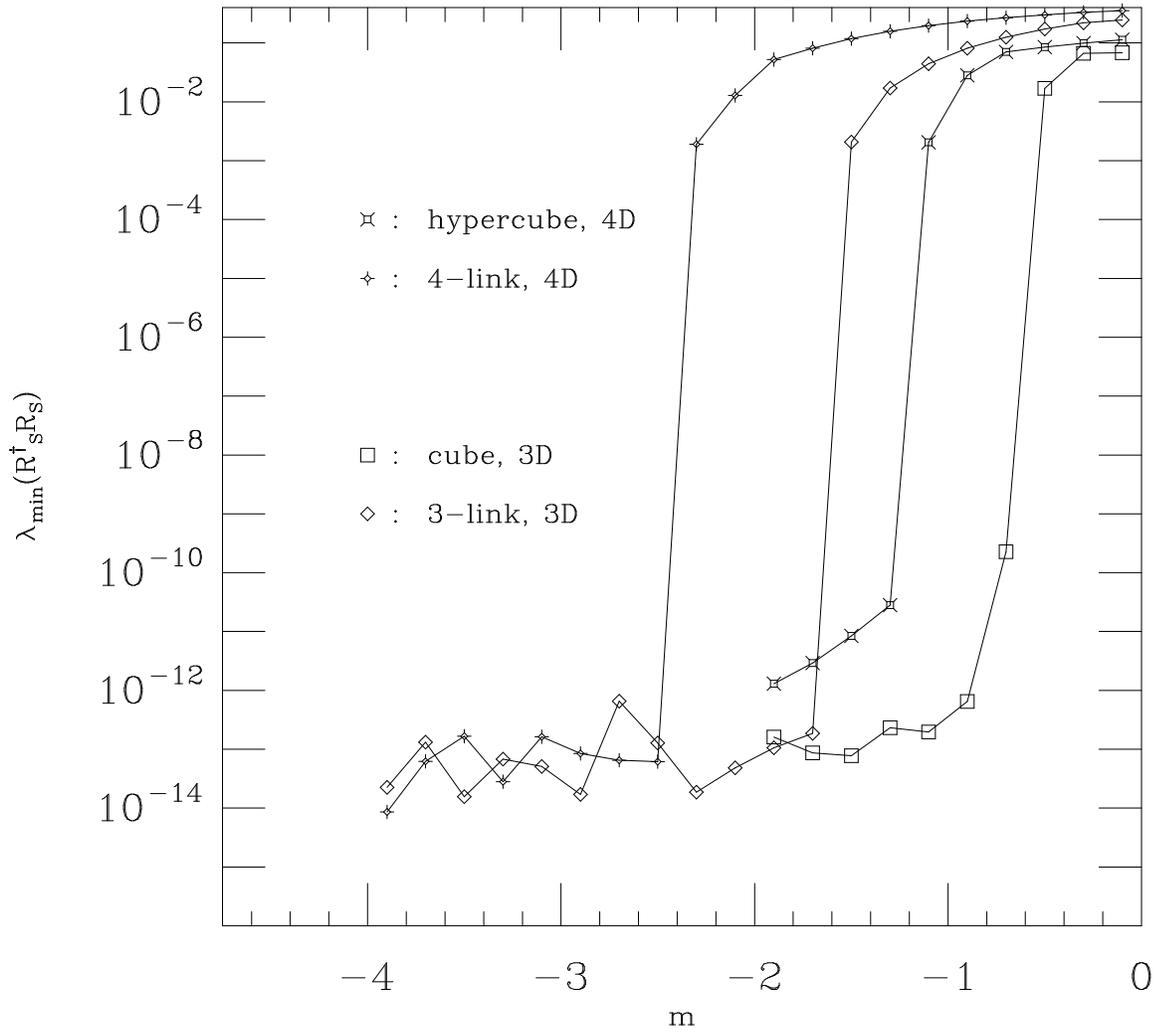}}}
\label{defect}
\end{figure}

The results of our numerical search are summarized in Fig.~\ref{defect}.
We plot there on logarithmic scale the minimum of the smallest
eigenvalue of $R_S^\dagger R_S$ as a function of the
Wilson mass parameter $m$, 
for four different clusters. The 3-link and 4-link clusters
in 3D and 4D mentioned in the figure 
contain 4 sites and 5 sites respectively and can
accommodate the fluxon configurations we studied analytically
before. The other two clusters we plot results for are the
cube in 3D and the hypercube in 4D. For example, we see in the figure 
that we can get zero modes for $m < -1.1$ in 4D if we consider
a cluster that makes up a hypercube. Note the sudden decrease
in $\lambda_{\rm min} (R_S^\dagger R_S )$ as a function of $m$;
it reflects a major change in the minimizing gauge configuration. 
Thus, for this type of cluster, $\lambda_{\rm min} (R_S^\dagger R_S )$
has a non-analytic dependence on the gauge background at some value
of $m$ located inside the range of interest $-2 < m <0$. Whatever
the ultimate role of this effect is in a full quenched simulation
of QCD is, it cannot be helpful to the hope of recovering 
a limit that can be described by a local effective Lagrangian of some
kind. 

\section{How to avoid zero modes}
We found that a cluster of SU(2) gauge fields 
as small as a hypercube can give rise to zero modes for Wilson
fermions at masses close enough to zero to be relevant to practical
simulations. These simulations can be with traditional fermions,
or with overlap fermions. In both cases the presence of fermionic 
zero modes is a problem for quenched simulations; this problem
has a practical solution in the overlap case, but they are still
costly to handle because they need to be individually identified
and projected out ~\cite{project}. 

Our construction directly and explicitly shows that the
number of fermionic zero modes is proportional to the lattice volume,
because the wave functions are explicitly known to decay exponentially,
and the clusters are small and can be well separated. The rate
of decay of the zero mode wave functions is given by the distance
of $m$ from the nearest even non-negative integer. The pure
gauge action for the clusters is exponentially suppressed and it
is possible that the density of the associated zero modes be so small
that in a practical simulation volume there would be a small chance
to even produce one such mode. On the other hand, we know that
small modes in numbers proportional to the lattice volume do occur
for quenched gauge field configurations. To be sure, we have not 
established that the particular clusters we investigate here
are responsible for all the small modes one sees in typical
QCD simulations. But, in principle, we have established that
there is a finite {\it density} of fermionic zero modes, and that
this density is exponentially suppressed in the gauge coupling constant,
consistent with numerical findings ~\cite{EHN}.
Clearly, in four dimensions, 
a crucial question is how this suppression relates to
the standard asymptotic freedom formula for a density of dimension
four. We hope to come back to this point in the future. 

It would be nice to be able to avoid these unphysical zero modes.
One can do that by changing the pure gauge action so that the
density of the zero modes be further reduced. This has been argued
to work to some extent ~\cite{EHN,iwasaki}. Another way would be,
in accordance with the exact bound ~\cite{bound}, to make the
local field strength close to zero. This could be achieved,
for example, by replacing all links by APE smeared ones ~\cite{degrand}.
However, this decouples the fermions from higher momentum modes
of the gauge field and makes it somewhat unclear whether we
are still dealing with a single scale problem, rather than
a two scale problem, where the fermions couple to gauge fields
only at the lower scale. 

Another possibility,
suggested in ~\cite{bound}, is to APE smear only the links that
enter into the Wilson mass term, but not those that are coupled
to the Dirac $\gamma$-matrices. In practice, implementing this
last alternative would roughly quadruple the time it takes to calculate
the action of the Wilson Dirac operator on a fermion field, because
the projector structure of the $1\pm\gamma_\mu$ terms is lost and 
the fermions now interact with two kinds of gauge fields.\footnote{We
thank Urs Heller for correcting our original erroneous statement
at this point.}

With our methods it is easy to obtain a rough estimate
of the effect on the zero  modes this latter possibility will have. 
We simply replace the links entering the Wilson
mass term by unity, leaving the links entering the other terms in
the Wilson Dirac operator free to change. This clearly is inconsistent
in terms of the entire gauge background, but is very easy to implement.
We find that this replacement of the Wilson mass term 
chases the fermionic zero modes away from
negative mass values too close to zero. 
This is an encouraging sign, and the idea
deserves further investigation.

\section {Conclusions}

In this paper we have devised a simple way to look for approximate
and exact fermionic zero modes on the lattice and for the gauge
configurations that bind them. The fermionic zero modes are bound
to local gauge configurations and are exponentially localized. In
a real quenched simulation they will come at a finite density making
the quenched approximation in principle unusable when traditional
fermions and a traditional gauge action are employed and one wants 
realistically light quarks to emerge in the continuum. 
Our method makes it obvious
that such fermion zero modes will occur with a finite
probability per unit Euclidean four volume.
We hope to use this technique in the reverse, namely, as a method 
to efficiently search for ways to ameliorate the practical problems
posed by fermionic almost zero modes in the overlap context. 

\acknowledgements

This research was 
supported in part by the DOE under grant \#
DE-FG05-96ER40559. F. B. thanks P. Sodano for discussions at earlier
stages of the project and gratefully acknowledges the Fondazione 
Angelo Della Riccia for financial support.

\end{document}